\pdfoutput=1
\documentclass[aps,prl,twocolumn,superscriptaddress,a4paper,
    floatfix]{revtex4}
\usepackage[latin1]{inputenc}
\usepackage{graphicx}
\usepackage{amsmath,amssymb}
\usepackage{hyperref}
\usepackage{datetime}

\begin{document}

\title{Direct Observation of Fragmentation in a Disordered, Strongly Interacting Fermi Gas}

\author{Sebastian Krinner}
\affiliation{Department of Physics, ETH Zurich, 8093 Zurich, Switzerland}
\author{David Stadler}
\affiliation{Department of Physics, ETH Zurich, 8093 Zurich, Switzerland}
\author{Jakob Meineke}
\affiliation{Department of Physics, ETH Zurich, 8093 Zurich, Switzerland}
\author{Jean-Philippe Brantut}
\affiliation{Department of Physics, ETH Zurich, 8093 Zurich, Switzerland}
\author{Tilman Esslinger}
\affiliation{Department of Physics, ETH Zurich, 8093 Zurich, Switzerland}

\date{\pdfdate}

\maketitle

{\bf 
Describing the behaviour of strongly interacting particles in the presence of disorder is among the most challenging problems in quantum many-body physics. The controlled setting of cold atom experiments provides a new avenue to address these challenges \cite{Shapiro:2012aa}, complementing studies in solid state physics, where a number of puzzling findings have emerged in experiments using superconducting thin films \cite{goldman_superconductor-insulator_1998,gantmakher_superconductorinsulator_2010}. Here we investigate a strongly interacting thin film of an atomic Fermi gas subject to a random potential. We use high-resolution in-situ imaging \cite{gemelke_situ_2009,Gericke:2008aa,Bakr:2009aa,Sherson:2010aa} to resolve the atomic density at the length scale of a single impurity, which would require scanning probe techniques in solid state physics \cite{sacepe_localization_2011}. This allows us to directly observe the fragmentation of the density profile and to extract its percolation properties. 
Transport measurements in a two-terminal configuration indicate that the fragmentation process is accompanied by a breakdown of superfluidity. Our results suggest that percolation of paired atoms is responsible for the loss of superfluidity, and that disorder is able to increase the binding energy of pairs.}

Two main scenarios have emerged, regarding the mechanism by which disorder can destroy conventional fermionic superfluidity in a thin film. In the first scenario, Cooper pairs are broken by disorder, and the unpaired particles get localised \cite{Finkelstein:1994aa}, whereas in the second scenario Cooper pairs are preserved in the presence of disorder and get themselves localised. The latter leads to the emergence of a bosonic insulator \cite{Larkin:1999aa} recently observed in-situ in an amorphous superconducting film \cite{sacepe_localization_2011}. 

To investigate this issue with cold atoms, we produce a cold atomic thin film by strongly confining a unitary Fermi gas along one direction. The resulting superfluid has a short coherence length \cite{bloch_many-body_2008}, a feature also encountered in exotic superconductors such as copper oxides \cite{gantmakher_superconductorinsulator_2010}. The thin film has a chemical potential $\mu\simeq1.9\,\hbar\omega_z= 0.55(7)\,\mu$K, where $\omega_z = 2\pi\cdot 6.1$\,kHz is the trap frequency along the tightly confined $z$ direction. This yields an interaction parameter $\rm{ln}(k_\mu a_{\rm{2D}}) = 1.5$ in the absence of disorder, corresponding to a strongly-interacting BCS-type regime \cite{frohlich_radio-frequency_2011,sommer_evolution_2012} (see Methods), with $a_{2D}$ denoting the two-dimensional scattering length and $k_\mu=\sqrt{2 m \mu/\hbar^2}$ the momentum associated with the chemical potential. 
We then add a controlled disorder in the film \cite{Shapiro:2012aa}, which introduces two new energy scales, the average disorder strength $\bar{V}$, and the correlation energy $E_\sigma = \hbar^2/m\sigma^2$, where $\sigma = 0.72(5)\,\mu$m is the disorder correlation length. We have $\mu > E_\sigma$, so that Anderson localisation of individual atoms at weak disorder should not occur \cite{Kuhn:2007aa,Shklovskii:2008aa}. However, we also have $E_\sigma = 2.4 \cdot E_b$, where $E_b = \hbar^2/ma_{\rm{2D}}^2$ is the binding energy of pairs, implying that disorder will exert different forces on the two paired constituents. In this regime, pairing is directly influenced by disorder and the question about the nature of the mechanism breaking superfluidity becomes relevant.

\begin{figure*}[htb]
    \includegraphics[width=0.95\textwidth]{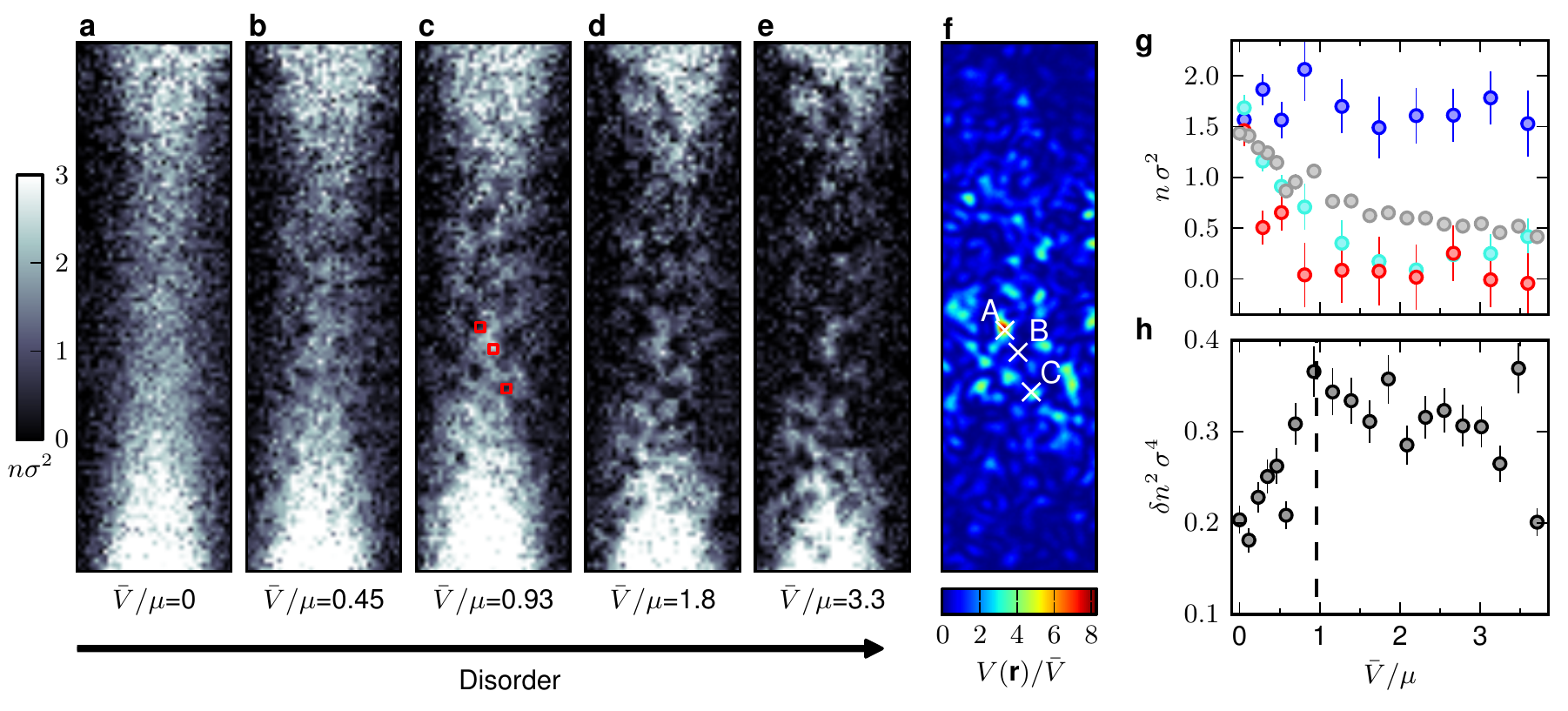}
    \caption{Evolution of the column density $n$ (in units of $\sigma^{-2}$) as the disorder strength is increased. a-e: High-resolution images of size $21\times72\,\mu$m of the in-situ density distribution in the channel for increasing $\bar{V}/\mu$. The saturated column density on top and bottom marks the beginning of the reservoirs, which extend far beyond the field of view. The systematic uncertainty in $\bar{V}/\mu$ is estimated to be 30\,\% (see Methods). f: Image of the projected speckle pattern. The density ripples, gradually appearing from figure a to e, can be matched one to one to bright (potential hills) and dark spots (potential valleys) in the image. 
    g: Local column density as a function of disorder strength for three specific points indicated in the potential landscape of panel f (point A: red, point B: blue, point C: cyan), each computed within a region of size $1.2 \times 1.2\,\mu$m marked as red squares in image c. The grey data points are the mean column density in the channel, computed in a central region of size $18\times 7\,\mu$m. h: Variance of the density computed in the same central region. The dashed line represents the theoretical percolation threshold for the potential seen by point-like pairs.}
    \label{fig:densityPics}
\end{figure*}

We first investigate the density distribution in the disordered potential using high-resolution imaging. Figure \ref{fig:densityPics} presents absorption images of the in-situ density for various disorder strengths. Figure \ref{fig:densityPics}a shows a clean film, smoothly connected on two sides to reservoirs, see Methods and \cite{brantut_conduction_2012-1} for details. For $\bar{V}/\mu=0.45$ (Fig. \ref{fig:densityPics}b), first density ripples appear. With increasing $\bar{V}/\mu$, they become more pronounced until $\bar{V}/\mu=1.8$ (Fig. \ref{fig:densityPics}d), where unpopulated regions occupy a significant fraction of the channel. At the largest disorder strength of $\bar{V}/\mu=3.3$ (Fig. \ref{fig:densityPics}e) the gas is composed of disconnected pockets separated by large empty regions. Figure \ref{fig:densityPics}f shows the potential landscape observed directly with our imaging system (see Supplementary Information). 
In figure \ref{fig:densityPics}g, the density at three distinct points (labeled A, B, C in figure \ref{fig:densityPics}f) is monitored as a function of disorder strength along the fragmentation process. Point A and C correspond to a large and a moderate potential hill and the local densities at these positions decrease correspondingly fast. In contrast, point B corresponds to a potential valley and its local density remains constant, suggesting that the superfluid persists locally at this point for all disorder strengths. The density averaged over the center part of the channel is shown for comparison in the same graph. It shows a smooth decrease with increasing disorder, due to the repulsive nature of the random potential. 

The appearance of density modulations can be quantified by extracting the variance of the density $\delta n^2$. It is presented in figure \ref{fig:densityPics}h as a function of disorder strength and shows a non-monotonous evolution: from zero disorder to $\bar{V}/\mu\thicksim 1$ density modulations increase quickly although the average density decreases in this interval. Having reached its maximum value at around $\bar{V}/\mu \sim 1$, the modulations slowly decrease for higher disorder, likely because the average density decreases. 

\begin{figure*}[htb]
    \includegraphics[width=0.95\textwidth]{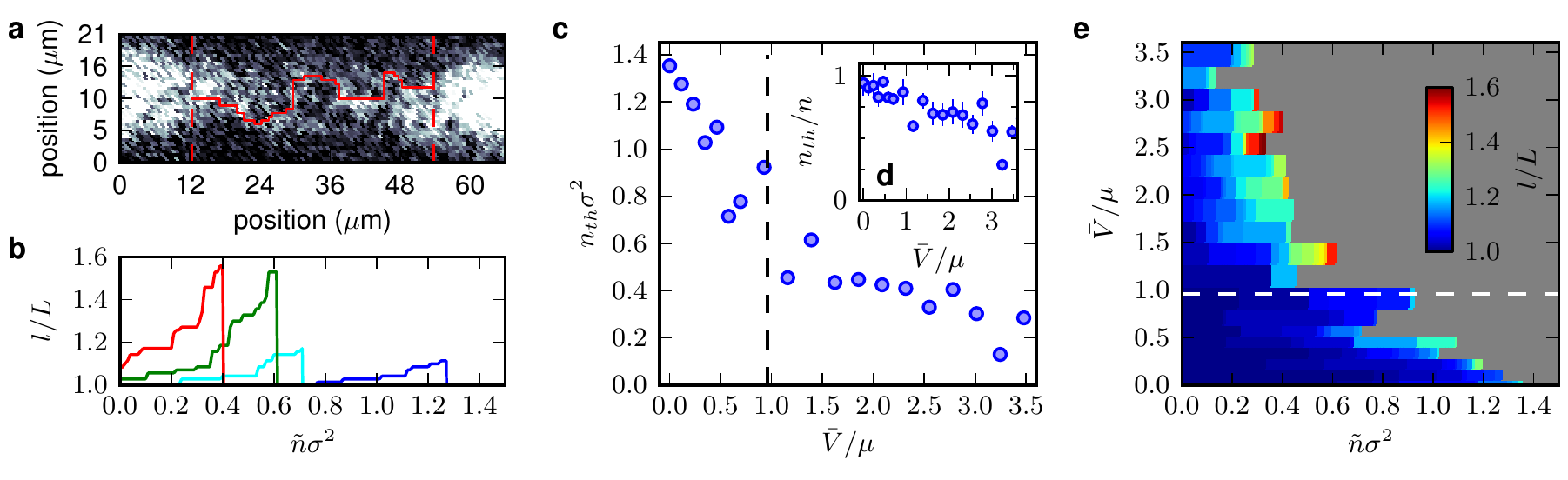}
    \caption{Percolation properties of the density distribution. a: Shortest connecting path between the two ends of the reservoirs (indicated by the red dashed lines) for $\bar{V}/\mu = 1.39$ and a density level of $\tilde{n}=0.6/\sigma^2$ close to the percolation threshold. The path displays strong deviations from a straight line and has a length of $1.5\cdot L$  b: Shortest path length as a function of density level $\tilde{n}$ (in units of $\sigma^{-2}$), for $\bar{V}/\mu = 0.12,\,0.58,\,1.39,\,2.78$ in blue, cyan, green, red, respectively. c: Percolation threshold as a function of disorder strength. d: Percolation threshold normalized to the average density as a function of disorder strength. Error bars represent the statistical error in the measurement of $n$. e: Map of the relative path length as a function of $\tilde{n}$ and $\bar{V}/\mu$. The dashed lines in c and e represent the theoretical percolation threshold of the potential seen by point-like pairs of atoms (see text).}
    \label{fig:pathLength}
\end{figure*}

The fragmentation process of the continuous density profile is naturally described in terms of percolation \cite{stauffer_introduction_1994}. We thus apply the tools of continuous percolation theory to the absorption images. For each disorder strength, we determine the length $l$ of the shortest possible connecting path from one reservoir to the other, along which the density $n$ always stays above a certain chosen density level $\tilde{n}$ (see Methods). The distance between the two ends of the reservoirs is $L=42\,\mu$m. A typical result is presented in figure \ref{fig:pathLength}a.

We evaluate the normalised path length $l/L$ as a function of $\tilde{n}$. The results are shown in figure \ref{fig:pathLength}b for different disorder strengths. Typically, $l/L$ remains close to one for $\tilde{n}$ much smaller than the mean density since the path remains close to a straight line. With increasing $\tilde{n}$, $l/L$ increases since regions of low density have to be circumvented. Beyond a critical threshold $n_{\mathrm{th}}$, no connecting path exists anymore. For large disorder strengths ($\bar{V}/\mu= 1.39,\,2.78$) $l/L$ reaches values as high as $1.6$ for $\tilde{n}$ close to $n_{th}$, limited by the extension of the cloud in the transverse direction. In contrast, for low disorder ($\bar{V}/\mu= 0.12,\,0.58$) the increase is limited to $l/L\sim 1.2$, which is comparable to the detection-noise induced increase without disorder (see Methods and Supplementary Information). 
These $l(\tilde{n})$ curves are typical for percolation transitions, and allow to unambiguously identify the percolation threshold of the density $n_{\mathrm{th}}$. The thresholds are plotted as a function of disorder strength in figure 2c. Two regimes can be identified: For $\bar{V}/\mu<1$, $n_{\mathrm{th}}$ shows a fast decrease since disorder grains of large amplitude quickly penetrate the film. For $\bar{V}/\mu>1$ the decrease is slowed down because most of the potential hills have already pierced the film. The transition coincides with the theoretically expected percolation threshold for tightly bound pairs, indicated by the vertical dashed line.

The ratio of the percolation threshold to the mean density is presented in figure \ref{fig:pathLength}d. It starts very close to one for weak disorder. Indeed, in a 2D continuous symmetric random profile the percolation threshold is equal to the average \cite{Zallen:1971aa}. For stronger disorder, this rule is violated and the threshold drops to values lower than the mean, indicating an increasing asymmetry in the density modulations due to the emergence of empty regions.

The path length $l/L$ as a function of disorder strength contains additional information. To show this we stack all $l(\tilde{n})/L$ curves as a function of disorder strength on the vertical axis, encoding $l/L$ in color. It is set to grey if no connecting path exists. The resulting 2D map is shown in figure \ref{fig:pathLength}e, manifesting the two regimes also in the path length: For $\bar{V} < \mu$ the maxima of $l/L$ reached at the percolation transitions remain moderate, whereas in the strongly disordered regime $\bar{V} > \mu$, large values of $l/L$ are encountered. The transition observed in figure \ref{fig:pathLength}c and \ref{fig:pathLength}e coincides with the maximum in the density modulations in figure \ref{fig:densityPics}h, confirming our interpretation of two regimes, smooth and fragmented. For comparison, we have reproduced the percolation analysis on a cloud subject to a homogeneous potential instead of the speckle pattern \cite{stadler_observing_2012}, finding only a single smooth regime (see Supplementary Information). 

\begin{figure}[ht]
    \includegraphics[width=0.45\textwidth]{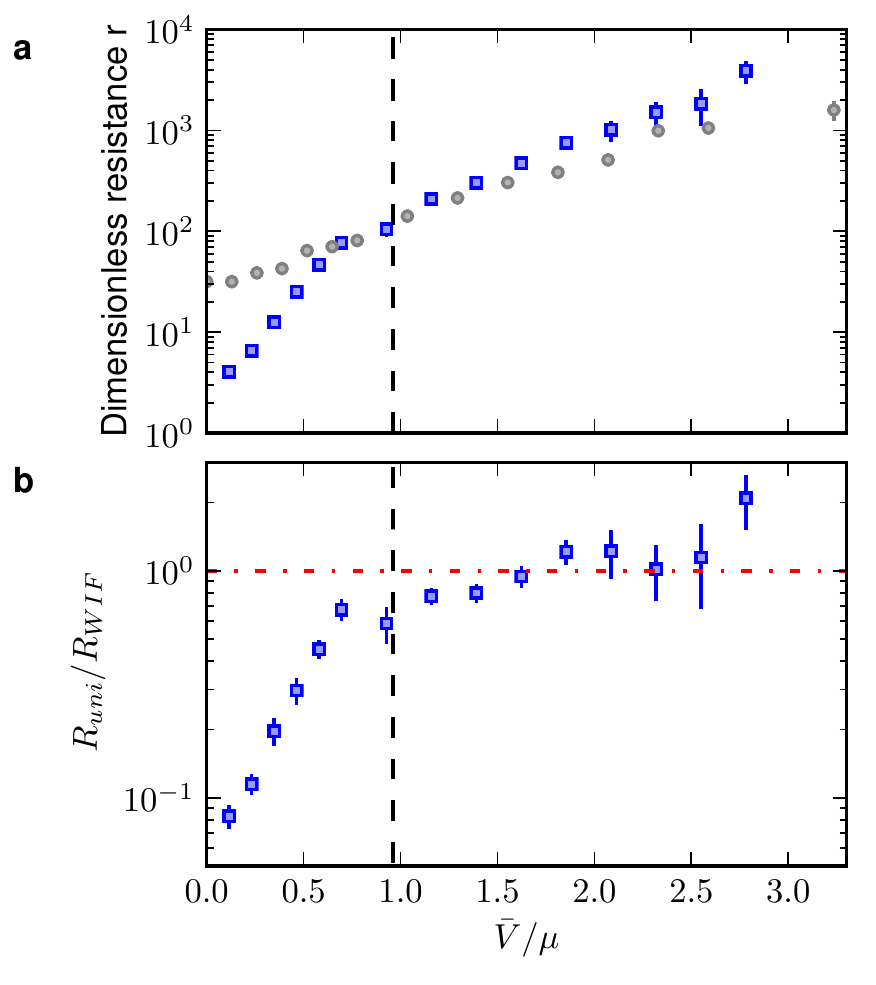}
    \caption{Resistance of the thin film. a: Dimensionless resistance as a function of disorder strength for the strongly interacting Fermi gas (blue squares) and for a weakly interacting Fermi gas (grey circles) for comparison. b: Ratio of absolute resistances of the strongly and weakly interacting Fermi gas. The dashed-dotted horizontal red line emphasizes that the ratio is close to one in the strongly disordered regime. The dashed vertical black line indicates the theoretical percolation threshold for the potential seen by tightly bound, point-like pairs.}
    \label{fig:transport}
\end{figure}

The evolution from smooth to fragmented density is accompanied by a clear change in the transport properties. We measure the dimensionless resistance $r$ of the thin film as a function of disorder strength (see Methods) \cite{stadler_observing_2012,Krinner:2013aa}. The results are presented in figure \ref{fig:transport}a and are similar to the case of a disordered bosonic superfluid \cite{Krinner:2013aa}. For the lowest disorder, the resistance is unmeasurably low, as it should be for a superfluid gas. The resistance increases very quickly until $\bar{V}/\mu\thicksim0.7$. Above this disorder strength the resistance increases more slowly. 
In order to disentangle the effects of strong interactions from single-particle effects, like simple atomic diffusion, we repeat the same experiment with a weakly interacting Fermi gas, for the same trapping potential, disorder configuration, and atom number. The resistances observed are shown in figure \ref{fig:transport}a as grey dots and show a smooth exponential evolution with disorder strength. For zero disorder, the resistance corresponds to the contact resistance of the ballistic channel \cite{brantut_conduction_2012-1}. 

We compare the transport properties of the two cases by evaluating the ratio of absolute resistances $R_{\mathrm{uni}}/R_{\mathrm{WIF}} = C_{\mathrm{WIF}}/C_{\mathrm{uni}}\cdot r_{\mathrm{uni}}/r_{\mathrm{WIF}}$  where $C_{\mathrm{uni}}$ ($C_{\mathrm{WIF}}$) is the compressibility of the reservoirs for the unitary (weakly interacting) Fermi gas. Assuming zero temperature and a harmonic trap for unitary and weakly interacting Fermi gases, one obtains $C_{\mathrm{WIF}}/C_{\mathrm{uni}}=\sqrt{\xi_{\rm{B}}}$, where $\xi_{\rm{B}}=0.38$ is the Bertsch parameter \cite{RevModPhys.80.1215}. Figure \ref{fig:transport}b shows the evolution of $R_{\mathrm{uni}}/R_{\mathrm{WIF}}$ with disorder strength. We observe a sharp decrease at $\bar{V}/\mu < 0.7$. Interestingly, $R_{\mathrm{uni}}/R_{\mathrm{WIF}}$ varies only very weakly and remains close to one in the strong disorder regime at $\bar{V}/\mu\gtrsim 1$. This suggests that beyond a certain level disorder dominates over interactions, even in a unitary Fermi gas.

The transition between the two regimes takes place close to the transition observed in the in-situ data. This suggests a percolation process driving the transition: For strong disorder, the superfluid fraction is localised and transport takes place as if the gas was normal. At the percolation threshold the superfluid islands start to connect and the resistance drops accordingly. The classical percolation threshold of the potential for a free atom of energy $\mu$ is reached at $\bar{V} = 1.92 \,\mu$ \cite{Weinrib:1982aa,Pezze:2011ab}. However, the potential seen by pairs is a more intricate problem and in the limit of tightly bound, point-like pairs it is equal to twice that for free atoms \cite{RevModPhys.80.1215}. In this extreme case, the percolation transition for pairs happens at $\bar{V} = 0.95 \,\mu$. This is indicated by the dashed, vertical lines in figures \ref{fig:densityPics}h, \ref{fig:pathLength}c, \ref{fig:pathLength}e, and \ref{fig:transport}. 
Our observations suggest that one can consider the pairs as tightly bound for $\bar{V}\gtrsim\mu$, and hence that disorder increases the binding energy of the pairs (see Supplementary Information for details). This case of tightly bound pairs being localised even while unpaired atoms are extended has been discussed in the context of the superconductor-insulator transition for a certain class of materials \cite{Lages:2000aa}.
  
Geometric percolation transitions play also a natural role in the disorder-induced superconductor-insulator transition in granular superconductors and may explain local \cite{sacepe_localization_2011} and global \cite{sherman_measurement_2012} tunneling spectroscopy measurements, as well as anomalous large magnetoresistance \cite{dubi_theory_2006} in homogeneously disordered amorphous films \cite{gantmakher_superconductorinsulator_2010}.

In future, the ability to locally observe the density distributions could be extended to a local measurement of the single-particle density of states using radio-frequency spectroscopy. This may allow to directly relate our findings to prominent models of disordered systems, such as a Bose-glass \cite{PhysRevB.37.325,fisher_boson_1989}, or a pseudo-gap phase \cite{Feigelman:2010aa}. 

We acknowledge fruitful discussions with Dima Shepelyanski, Gabriel Lemari\'e, Thierry Giamarchi, Dan Shahar, Vijay Shenoy, Vincent Josse, Antoine Georges, Corina Kollath, Charles Grenier and Sebastiano Pilati. We acknowledge financing from NCCR MaNEP and QSIT, the ERC Projects SQMS and SIQS, the FP7 Project NAME-QUAM, the EU through the Collaborative Project 600645 SIQS, and ETHZ. J.P.B. is supported by the EU through a Marie Curie Fellowship.

\section*{Methods}

\subsection*{Experimental setup} Our experiment uses the system described in \cite{brantut_conduction_2012-1, stadler_observing_2012}. In brief, a degenerate Fermi gas of about $10^5$ atoms of $^6$Li in a balanced mixture of the two lowest hyperfine states is produced by forced evaporation in an optical dipole trap. The experiments are performed in a homogeneous magnetic field of 834\,G, where the gas is in the unitary regime and resides in a superfluid state. A tightly confining channel with a central trap frequency of $\omega_z=2\pi\cdot 6.1$\,kHz is imprinted at the center of the cigar-shaped cloud, creating two atomic reservoirs, connected by a quasi-2D channel. Two microscope objectives are used to address the atoms in the constriction \cite{b_zimmermann_and_t_muller_and_j_meineke_and_t_esslinger_and_h_moritz_high-resolution_2011}. 

The disordered potential is generated optically: we use a speckle pattern, produced by a laser at 532\,nm, and project it on the constriction along the $z$-axis through the upper objective. It can be imaged by the lower microscope objective, allowing for a precise characterization of the disordered potential. The pattern has a gaussian envelope with a waist of 35\,$\mu$m.
We characterize its strength by the ac Stark shift at the maximum value $\bar{V}$ of the envelope. The uncertainty in $\bar{V}$ amounts to 20\% and is due to optical power measurement. The correlation length $\sigma$ of the speckle pattern is defined as the $1/\sqrt{e}$\,-\,radius of a gaussian fit to the autocorrelation function of the speckle pattern and yields $0.72\,\mu$m.

Starting from a strongly interacting gas in the channel, we switch on the disorder to a variable strength and wait for 150 ms for thermalization with the reservoirs. The atoms are then illuminated by a 4\,$\mu$s pulse of resonant light with an intensity of about $0.1$\,$I_{\mathrm{sat}}$, where $I_{\mathrm{sat}}$ is the saturation intensity of the transition, and the absorption pattern is registered on an EMCCD camera. We average typically 20 of those pictures to reduce noise, leading to the images shown in figure \ref{fig:densityPics} and analysed in figure \ref{fig:pathLength}.

\subsection*{Fermi gas in a random potential}

At low temperature, the unitary Fermi gas, strongly confined along one direction, has two energy scales: The chemical potential $\mu= 0.55(7)\,\mu$K, obtained by assuming a harmonically trapped zero temperature unitary Fermi gas, and the binding energy of pairs $E_b$. Since the film is connected to macroscopic reservoirs that are not affected by disorder, the chemical potential in the channel is fixed. With $\mu\simeq1.9\,\hbar\omega_z$, where $\omega_z = 2\pi\cdot 6.1$\,kHz is the trap frequency along the tightly confined direction, the gas is in the confinement dominated regime \cite{petrov_interatomic_2001}. 
In a gas tightly confined in one direction, we have $E_b=0.24$\,$\hbar \omega_z$ \cite{frohlich_radio-frequency_2011,sommer_evolution_2012}. This energy scale defines the 2D scattering length $a_{\mathrm{2D}} =\hbar/\sqrt{m E_b}$, $m$ denoting the mass of lithium atoms, which is also the typical size of a pair of atoms. We parametrize the interaction strength by $\rm{ln}(k_\mu a_{\rm{2D}})$, where $k_\mu=\frac{\sqrt{2 m \mu}}{\hbar}$ is the momentum of a particle at energy $\mu$, with BEC or BCS regimes corresponding to $\rm{ln}(k_\mu a_{\rm{2D}})$ negative or positive. In our experiment, without disorder we have $\rm{ln}(k_\mu a_{\rm{2D}}) = 1.5$. 

\subsection{Percolation analysis}
The shortest possible connecting path between the two reservoirs for a given disorder strength and density level $\tilde{n}$ is determined as follows: in a window of size $42\times24\mu$m, comprising the full film and overlapping with the ends of the reservoirs (see figure \ref{fig:pathLength}a), we set all pixels (size $0.6\times0.6\,\mu$m) with density $n\geq\tilde{n}$ to one and all pixels with $n<\tilde{n}$ to zero. We then run the burning algorithm \cite{stauffer_introduction_1994} on this binary grid, retaining the length $l$ of the shortest possible path. Due to the inherent detection noise the recorded density profile is not perfectly homogenous even at zero disorder potential and therefore also leads to a moderate increase in path length when $\tilde{n}$ is increased (see Supplementary Information).

\subsection{Resistance measurement}

The transport experiment is performed using the technique presented in \cite{brantut_conduction_2012-1}. The disordered film is smoothly attached on two sides to macroscopic atomic reservoirs, which have identical geometry and atom numbers $N_1$ and $N_2$. During preparation a constant magnetic field gradient of $2.5$\,mT/m along the long axis (transport axis) of the trap is applied in order to shift the trap with respect to the center of the channel. We restore the symmetry of the trapping potential by ramping the magnetic field gradient to zero within 10\,ms. This procedure leads to a low temperature unitary superfluid in each reservoir with a relative imbalance $(N_1 - N_2)/(N_1+N_2)=0.3$, corresponding to a chemical potential bias of $\Delta\mu\simeq0.1\mu$. As a response to this bias an atomic current sets in. The starting point of the transport measurement is defined by the endpoint of the ramp-down of the magnetic field gradient.

We fit the time evolution of $(N_1 - N_2)/(N_1+N_2)$ with a decreasing exponential, providing a decay constant $\tau$. This fit accurately reproduces the data for $\bar{V}/\mu > 0.2$. In this case, the time evolution is analogous to the discharge of a capacitor through a resistor. The compressibility of the reservoirs, which plays the role of the capacity of the capacitor, is not affected by the disorder and thus remains constant. Hence, the evolution of $\tau$ with disorder strength directly reflects the evolution of the resistance of the film with disorder strength. We normalize $\tau$ by the trap frequency along the transport axis in the absence of disorder and channel to obtain the dimensionless resistance $r$ \cite{stadler_observing_2012}. Error bars correspond to fit errors.

The weakly interacting Fermi gas is prepared at a magnetic field of $475\,G$, where the scattering length is $-100\,a_0$, and $a_0$ is the Bohr radius. This results in a weakly interacting Fermi gas (WIF) of about the same atom number at a temperature of 0.25(5)\,$T_F$, $T_F=1.0(2)\,\mu$K denoting the Fermi temperature in this case.

\bibliographystyle{naturemag}

\clearpage
\onecolumngrid
\section{Supplementary Information} 
\renewcommand{\theequation}{S\arabic{equation}}
\renewcommand{\thefigure}{S\arabic{figure}}
\renewcommand*{\citenumfont}[1]{S#1}
\renewcommand*{\bibnumfmt}[1]{[S#1]}

\subsection*{Observation of the potential landscape at the single impurity level}

\begin{figure}[htbp]
\def \thefigure{S1}
\begin{center}
\includegraphics[width=0.5\textwidth]{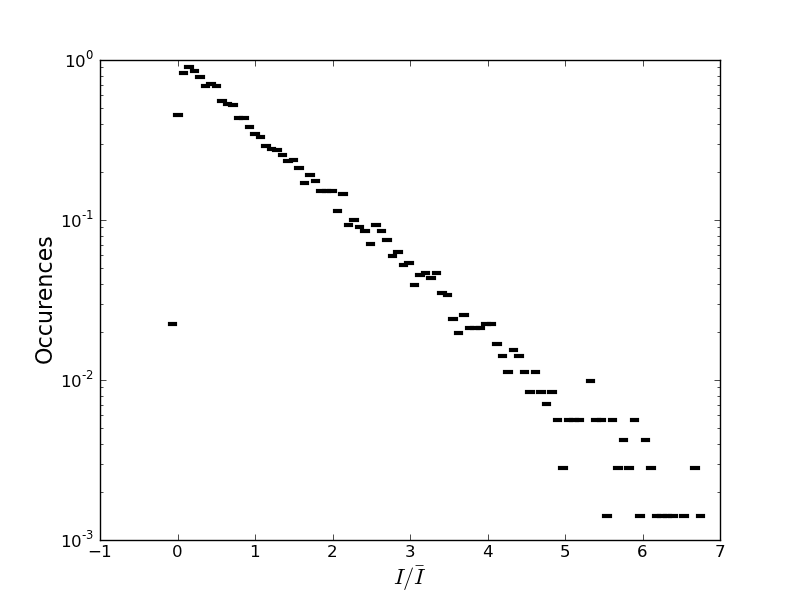}
\caption{Normalized histogram of the intensity distribution of the speckle pattern observed through the imaging system. An exponential shape of the distribution is expected for fully resolved speckle, and the rounding observed at low values of $I/\bar{I}$ can be used to estimate the finite resolution effects.}
\label{fig:hist}
\end{center}
\end{figure}

Our experimental system comprises two identical microscope objectives facing each other. One of them is used to create the fine grained speckle pattern, and the other is used to image both the intensity distribution of the speckle and the density of the atoms with negligible chromatic aberrations \cite{Zimmermann:2011aa}. We now analyse the pictures of the speckle pattern to quantify the ratio of optical resolution to the correlation length of the disorder. To this end, we look at the central part of the speckle pattern (30x30\,$\mu$m) where the atoms reside, and plot the histogram of the intensity distribution in all pixels of size $300$x$300$\,nm. The normalised histogram is presented in figure \ref{fig:hist}, as a function of the relative intensity $I/\bar{I}$ where $\bar{I}$ is the mean intensity.

We observe an exponential distribution over three orders of magnitude for the intensity, which is predicted for the fully developed speckle pattern \cite{Goodman:2007aa}. The effects of finite resolution of the optical system manifest themselves in the rounding of the intensity distribution close to zero: the blurring of the speckle pattern strongly reduces the probability to observe very low intensity regions. 

By supposing that the rounding observed in figure \ref{fig:hist} is entirely due to finite resolution effects, we obtain an upper bound of the ratio of the optical resolution area to the correlation area. We use the approximate probability distribution of integrated speckle given in the literature \cite{Goodman:2007aa} and evaluate the amount of blurring from the position and height of the maximum in the histogram. This yields a ratio of resolution area  to correlation area of $0.11(2)$, where the uncertainty comes from finite sample-size noise in the histogram. This low value confirms that our imaging system is capable to resolve single modulations of the potential, which we are able to relate one to one to modulations in the density profile (figure 1 in main text).

The correlation length of the speckle pattern is obtained by fitting a gaussian to the autocorrelation function of the speckle pattern, like in our previous works (for example \cite{Brantut:2012aa}). This analysis neglects the effects of finite optical resolution. This correlation length $\sigma=0.72\,\mu$m and the numerical aperture $\mathrm{NA}=0.53$ of our imaging system provide another possibility to calculate the ratio of optical resolution area $\pi(\lambda/2\mathrm{NA})^2$ to correlation area $\pi\sigma^2$, yielding an expected value of 0.12, in agreement with the above estimate.

\subsection*{Scattering and Feshbach resonances in a random potential}

Consider two atoms confined to two dimensions in the presence of a random potential, with average amplitude $\bar{V}$ and correlation length $\sigma$. The classical frequency scale for the oscillations in a local minimum of the random potential is $\Omega=\frac{1}{\sigma} \sqrt{ \frac{\bar{V}}{m}}$. In analogy with optical lattices, we can write $\hbar \Omega = \sqrt{\bar{V} E_{\sigma}}$ with $E_{\sigma} = \frac{\hbar^2}{m \sigma^2}$ the correlation energy, which is the counterpart of the recoil energy in an optical lattice \cite{Morsch:2006aa}.

\begin{figure}[htbp]
\def \thefigure{S2}
\begin{center}
\includegraphics[width=0.5\textwidth]{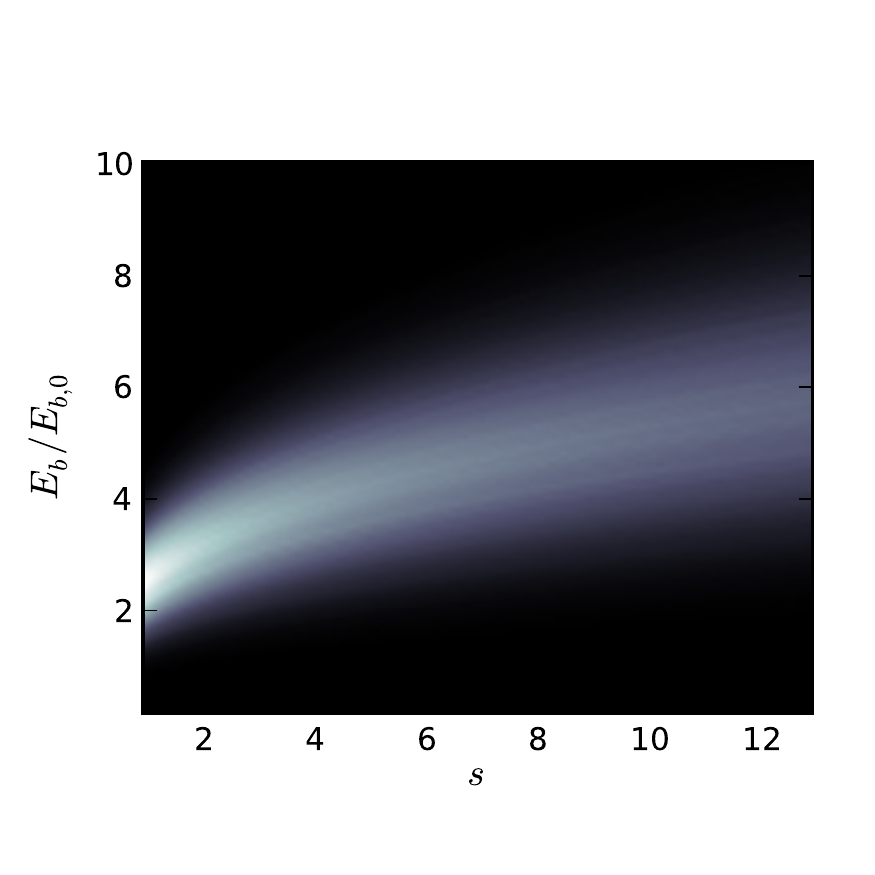}
\caption{Statistical distribution of binding energies in a thin film of cold Fermions subject to a classical speckle potential, normalised by the binding energy in the absence of disorder. Each vertical cut is a normalised histogram of occurrences of binding energies.}
\label{fig:s1}
\end{center}
\end{figure}

The presence of disorder modifies the boundary conditions for the scattering problem, by localising the atoms. We now illustrate the effect of disorder on scattering with the case of a classical speckle potential $\bar{V} \gg E_{\sigma}$. Here atoms get increasingly localised in local potential minima, where the trapping potential is approximately harmonic. Let $\omega_z$ be the trap frequency along the direction perpendicular to the plane of the film, and $\omega_1,\omega_2$ the local trap frequencies in the plane. 
On the Feshbach resonance, where the 3D scattering length diverges, the two-body binding energy of the two atoms is $E_b = \hbar (\omega_z \omega_1 \omega_2)^{1/3}$ \cite{Busch:1998aa}. With increasing disorder strength $\bar{V}$, the trap frequencies in the plane increase and the binding energy increases as well. In the absence of disorder, the binding energy is equal to $E_{b,0} = 0.24\, \hbar \omega_z$ \cite{Petrov:2001aa,Frohlich:2011aa,Sommer:2012aa}. In a situation where the speckle is isotropic,  we write $\sqrt{\omega_1 \omega_2} = \omega_{\parallel} = \alpha \Omega$, defining $\alpha$ as the in-plane trap frequency in units of $\Omega$. The binding energy of pairs then reads
\begin{equation}
\frac{E_b}{E_{b,0}} = \frac{1}{0.24} \left( \frac{E_{\sigma}}{\hbar \omega_z} \right)^{2/3} s^{1/3} \alpha^{2/3},
\label{eq:binding}
\end{equation}
where $s=\bar{V}/E_{\sigma}$ is the average disorder strength normalised to the correlation energy.
We now estimate explicitly this effect for our experiment, where $\omega_z = 6.1\,$kHz, and $E_{\sigma} = \mathrm{h} \cdot 3.2\,$kHz. We have numerically determined the statistical distribution of $\alpha$ in speckle patterns. The distribution of binding energies is then computed from \ref{eq:binding}. The results are presented in figure \ref{fig:s1} as a function of $s$. The maximum value shown, $s=13$, corresponds to the maximum value of $\bar{V}/\mu\simeq 3.5$ explored in the experiment. As expected from \ref{eq:binding} the mean of the binding energy grows proportional to $s^{1/3}$. The relative width of the distribution $\delta E_b/E_b\simeq1$ is independent of $s$. Already at the validity limit $s\gtrsim 1$ of this classial regime, $E_b$ is larger than twice the binding energy of the clean system, showing that the effects of disorder dominate the pure effect of the confinement along $z$.

Hence, we conclude that disorder tends to increase the binding energy of pairs, despite the fact that the pairs in a clean film are only loosly bound. In the BEC-BCS picture, this means that a sufficiently strong disorder will always drive the gas towards the BEC regime, and that the natural candidate for the strongly disordered gas is a Bose-glass \cite{Giamarchi:1988aa,Fisher:1989aa}.

Interestingly, disorder also introduces random fluctuations of the binding energy, reflected in the broadening of the binding energy distributions. Extending the usual definition of the 2D scattering length $a_{2D} = \sqrt{\hbar^2/mE_b}$ to the disordered case, we are led to the conclusion that the gas acquires randomness in the two-body scattering itself. Models with random interactions present some differences with respect to the usual case  of random potential background \cite{Gimperlein:2005aa}. They have been introduced in nuclear \cite{Johnson:1998aa,Papenbrock:2002aa} and mesoscopic physics \cite{Jacquod:2000aa}, and cold atoms could allow their experimental investigation in a novel environment. 

We have discussed the case of the classical speckle for simplicity, but the picture should remain qualitatively valid if atoms get localised by quantum interference rather than by classical trapping. In this case, a localisation energy $E_{\xi} = \frac{\hbar^2}{m \xi^2}$, where $\xi$ is the localisation length, will play the role of the oscillation frequency in one minima, and at constant $s$ we expect an increase of the binding energy like $E_{\xi}^{2/3}$. 

Also, these effects where computed at the point where the scattering length in three dimensions diverges, but the qualitative effect of disorder on binding should remain valid for finite but large scattering length, comparable or larger than the correlation length of the disorder.

\subsection*{Percolation analysis for a homogenous potential}

We have applied the percolation analysis leading to figure 2e of the main text to the situation where the disordered potential is replaced by a homogenous repulsive potential with a gaussian envolope of waists $w_x=w_y=18\mu$m \cite{stadler_observing_2012}. Here the atom number was $7\times 10^4$ and the trapping frequency of the quasi-2D confinement was 2.9\,kHz. The distance $L$ within which the algorithm for finding the shortest path $l$ was run, was reduced to $L=24\mu$m to take into account the smaller waist of the homogenous gate beam as compared to the speckle beam. 
\begin{figure}[htbp]
\def \thefigure{S3}
\begin{center}
\includegraphics[width=0.5\textwidth]{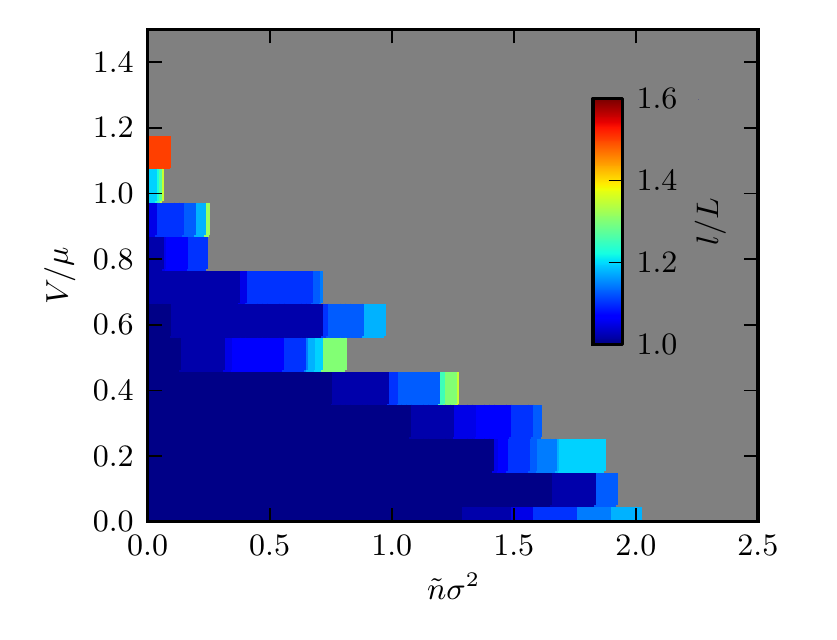}
\caption{Map of normalised path length as a function of disorder strength (vertical axis) and density level (horizontal axis).}
\label{fig:gate}
\end{center}
\end{figure}
Figure \ref{fig:gate} shows a false-color representation of the normalised path length $l/L$ in complete analogy to figure 2e of the main text. For the density level plotted on the x-axis we stick to the same units as in the paper, i.e. atoms per correlation area. The potential amplitude $V$, plotted normalized to $\mu$ on the vertical axis, refers to the maximum of the repulsive gate potential. With increasing gate power the threshold density $n_{th}$, marked by the interface of coloured and grey regions, decreases quickly, similar to the initial fast reduction observed in the disordered case. This is because the repulsive gate beam continously reduces the density in the center of the channel. 
For $V\gtrsim \mu$ there is no connecting path for a non-zero $\tilde{n}$ any more because the density in the center of the channel has been entirely depleted. This is in strong contrast to the disordered case of figure 2e in the main text, where an additional regime of small $n_{th}$ and large path lengths is observed for $\bar{V}\gtrsim \mu$. Note also that the normalised path lengths at the thresholds may increase to values slightly above the expected value of one for a perfectly homogenous density profile because the recorded density profiles are not noise-free. This residual increase up to values of $1.2\times L$ is the same in the case of the homogenous gate beam and in the case of the smooth regime of the disordered case.

\bibliographystyle{naturemag}

\begin{thebibliography}{10}
\expandafter\ifx\csname url\endcsname\relax
  \def\url#1{\texttt{#1}}\fi
\expandafter\ifx\csname urlprefix\endcsname\relax\def\urlprefix{URL }\fi
\providecommand{\bibinfo}[2]{#2}
\providecommand{\eprint}[2][]{\url{#2}}

\bibitem{Shapiro:2012aa}
\bibinfo{author}{Shapiro, B.}
\newblock \bibinfo{title}{Cold atoms in the presence of disorder}.
\newblock \emph{\bibinfo{journal}{Journal of Physics A: Mathematical and
  Theoretical}} \textbf{\bibinfo{volume}{45}}, \bibinfo{pages}{143001}
  (\bibinfo{year}{2012}).

\bibitem{goldman_superconductor-insulator_1998}
\bibinfo{author}{Goldman, A.~M.} \& \bibinfo{author}{Markovic, N.}
\newblock \bibinfo{title}{{Superconductor-Insulator} transitions in the
  {Two-Dimensional} limit}.
\newblock \emph{\bibinfo{journal}{Physics Today}}
  \textbf{\bibinfo{volume}{51}}, \bibinfo{pages}{39--44}
  (\bibinfo{year}{1998}).

\bibitem{gantmakher_superconductorinsulator_2010}
\bibinfo{author}{Gantmakher, V.~F.} \& \bibinfo{author}{Dolgopolov, V.~T.}
\newblock \bibinfo{title}{Superconductor-insulator quantum phase transition}.
\newblock \emph{\bibinfo{journal}{Physics-Uspekhi}}
  \textbf{\bibinfo{volume}{53}}, \bibinfo{pages}{1--49} (\bibinfo{year}{2010}).

\bibitem{gemelke_situ_2009}
\bibinfo{author}{Gemelke, N.}, \bibinfo{author}{Zhang, X.},
  \bibinfo{author}{Hung, C.-L.} \& \bibinfo{author}{Chin, C.}
\newblock \bibinfo{title}{In situ observation of incompressible mott-insulating
  domains in ultracold atomic gases}.
\newblock \emph{\bibinfo{journal}{Nature}} \textbf{\bibinfo{volume}{460}},
  \bibinfo{pages}{995--998} (\bibinfo{year}{2009}).

\bibitem{Gericke:2008aa}
\bibinfo{author}{Gericke, T.}, \bibinfo{author}{Wurtz, P.},
  \bibinfo{author}{Reitz, D.}, \bibinfo{author}{Langen, T.} \&
  \bibinfo{author}{Ott, H.}
\newblock \bibinfo{title}{High-resolution scanning electron microscopy of an
  ultracold quantum gas}.
\newblock \emph{\bibinfo{journal}{Nat Phys}} \textbf{\bibinfo{volume}{4}},
  \bibinfo{pages}{949--953} (\bibinfo{year}{2008}).

\bibitem{Bakr:2009aa}
\bibinfo{author}{{Bakr}, W.~S.}, \bibinfo{author}{{Gillen}, J.~I.},
  \bibinfo{author}{{Peng}, A.}, \bibinfo{author}{{F{\"o}lling}, S.} \&
  \bibinfo{author}{{Greiner}, M.}
\newblock \bibinfo{title}{{A quantum gas microscope for detecting single atoms
  in a Hubbard-regime optical lattice}}.
\newblock \emph{\bibinfo{journal}{Nature}} \textbf{\bibinfo{volume}{462}},
  \bibinfo{pages}{74--77} (\bibinfo{year}{2009}).

\bibitem{Sherson:2010aa}
\bibinfo{author}{{Sherson}, J.~F.} \emph{et~al.}
\newblock \bibinfo{title}{{Single-atom-resolved fluorescence imaging of an
  atomic Mott insulator}}.
\newblock \emph{\bibinfo{journal}{Nature}} \textbf{\bibinfo{volume}{467}},
  \bibinfo{pages}{68--72} (\bibinfo{year}{2010}).

\bibitem{sacepe_localization_2011}
\bibinfo{author}{Sac{\'e}p{\'e}, B.} \emph{et~al.}
\newblock \bibinfo{title}{Localization of preformed cooper pairs in disordered
  superconductors}.
\newblock \emph{\bibinfo{journal}{Nature Physics}}
  \textbf{\bibinfo{volume}{7}}, \bibinfo{pages}{239--244}
  (\bibinfo{year}{2011}).

\bibitem{Finkelstein:1994aa}
\bibinfo{author}{Finkel'stein, A.~M.}
\newblock \bibinfo{title}{Suppression of superconductivity in homogeneously
  disordered systems}.
\newblock \emph{\bibinfo{journal}{Physica B: Condensed Matter}}
  \textbf{\bibinfo{volume}{197}}, \bibinfo{pages}{636--648}
  (\bibinfo{year}{1994}).

\bibitem{Larkin:1999aa}
\bibinfo{author}{Larkin, A.}
\newblock \bibinfo{title}{Superconductor-insulator transitions in films and
  bulk materials}.
\newblock \emph{\bibinfo{journal}{Annalen der Physik}}
  \textbf{\bibinfo{volume}{8}}, \bibinfo{pages}{785--794}
  (\bibinfo{year}{1999}).

\bibitem{bloch_many-body_2008}
\bibinfo{author}{Bloch, I.}, \bibinfo{author}{Dalibard, J.} \&
  \bibinfo{author}{Zwerger, W.}
\newblock \bibinfo{title}{Many-body physics with ultracold gases}.
\newblock \emph{\bibinfo{journal}{Reviews of Modern Physics}}
  \textbf{\bibinfo{volume}{80}}, \bibinfo{pages}{885--964}
  (\bibinfo{year}{2008}).

\bibitem{frohlich_radio-frequency_2011}
\bibinfo{author}{Fr{\"o}hlich, B.} \emph{et~al.}
\newblock \bibinfo{title}{Radio-frequency spectroscopy of a strongly
  interacting two-dimensional fermi gas}.
\newblock \emph{\bibinfo{journal}{Physical Review Letters}}
  \textbf{\bibinfo{volume}{106}}, \bibinfo{pages}{105301}
  (\bibinfo{year}{2011}).

\bibitem{sommer_evolution_2012}
\bibinfo{author}{Sommer, A.~T.}, \bibinfo{author}{Cheuk, L.~W.},
  \bibinfo{author}{Ku, M. J.~H.}, \bibinfo{author}{Bakr, W.~S.} \&
  \bibinfo{author}{Zwierlein, M.~W.}
\newblock \bibinfo{title}{Evolution of fermion pairing from three to two
  dimensions}.
\newblock \emph{\bibinfo{journal}{Physical Review Letters}}
  \textbf{\bibinfo{volume}{108}}, \bibinfo{pages}{045302}
  (\bibinfo{year}{2012}).

\bibitem{Kuhn:2007aa}
\bibinfo{author}{{Kuhn}, R.~C.}, \bibinfo{author}{{Sigwarth}, O.},
  \bibinfo{author}{{Miniatura}, C.}, \bibinfo{author}{{Delande}, D.} \&
  \bibinfo{author}{{M{\"u}ller}, C.~A.}
\newblock \bibinfo{title}{{Coherent matter wave transport in speckle
  potentials}}.
\newblock \emph{\bibinfo{journal}{New Journal of Physics}}
  \textbf{\bibinfo{volume}{9}}, \bibinfo{pages}{161--+} (\bibinfo{year}{2007}).

\bibitem{Shklovskii:2008aa}
\bibinfo{author}{Shklovskii, B.}
\newblock \bibinfo{title}{Superfluid-insulator transition in ``dirty''
  ultracold fermi gas}.
\newblock \emph{\bibinfo{journal}{Semiconductors}}
  \textbf{\bibinfo{volume}{42}}, \bibinfo{pages}{909--913}
  (\bibinfo{year}{2008}).

\bibitem{brantut_conduction_2012-1}
\bibinfo{author}{Brantut, J.-P.}, \bibinfo{author}{Meineke, J.},
  \bibinfo{author}{Stadler, D.}, \bibinfo{author}{Krinner, S.} \&
  \bibinfo{author}{Esslinger, T.}
\newblock \bibinfo{title}{Conduction of ultracold fermions through a mesoscopic
  channel}.
\newblock \emph{\bibinfo{journal}{Science}} \textbf{\bibinfo{volume}{337}},
  \bibinfo{pages}{1069--1071} (\bibinfo{year}{2012}).

\bibitem{stauffer_introduction_1994}
\bibinfo{author}{Stauffer, D.} \& \bibinfo{author}{Aharony, A.}
\newblock \emph{\bibinfo{title}{Introduction To Percolation Theory}}
  (\bibinfo{publisher}{Taylor \& Francis}, \bibinfo{year}{1994}).

\bibitem{Zallen:1971aa}
\bibinfo{author}{Zallen, R.} \& \bibinfo{author}{Scher, H.}
\newblock \bibinfo{title}{Percolation on a continuum and the
  localization-delocalization transition in amorphous semiconductors}.
\newblock \emph{\bibinfo{journal}{Phys. Rev. B}} \textbf{\bibinfo{volume}{4}},
  \bibinfo{pages}{4471--4479} (\bibinfo{year}{1971}).

\bibitem{stadler_observing_2012}
\bibinfo{author}{Stadler, D.}, \bibinfo{author}{Krinner, S.},
  \bibinfo{author}{Meineke, J.}, \bibinfo{author}{Brantut, J.-P.} \&
  \bibinfo{author}{Esslinger, T.}
\newblock \bibinfo{title}{Observing the drop of resistance in the flow of a
  superfluid fermi gas}.
\newblock \emph{\bibinfo{journal}{Nature}} \textbf{\bibinfo{volume}{491}},
  \bibinfo{pages}{736--739} (\bibinfo{year}{2012}).

\bibitem{Krinner:2013aa}
\bibinfo{author}{Krinner, S.}, \bibinfo{author}{Stadler, D.},
  \bibinfo{author}{Meineke, J.}, \bibinfo{author}{Brantut, J.-P.} \&
  \bibinfo{author}{Esslinger, T.}
\newblock \bibinfo{title}{Superfluidity with disorder in a thin film of quantum
  gas}.
\newblock \emph{\bibinfo{journal}{Phys. Rev. Lett.}}
  \textbf{\bibinfo{volume}{110}}, \bibinfo{pages}{100601}
  (\bibinfo{year}{2013}).

\bibitem{RevModPhys.80.1215}
\bibinfo{author}{Giorgini, S.}, \bibinfo{author}{Pitaevskii, L.~P.} \&
  \bibinfo{author}{Stringari, S.}
\newblock \bibinfo{title}{Theory of ultracold atomic fermi gases}.
\newblock \emph{\bibinfo{journal}{Rev. Mod. Phys.}}
  \textbf{\bibinfo{volume}{80}}, \bibinfo{pages}{1215--1274}
  (\bibinfo{year}{2008}).

\bibitem{Weinrib:1982aa}
\bibinfo{author}{Weinrib, A.}
\newblock \bibinfo{title}{Percolation threshold of a two-dimensional continuum
  system}.
\newblock \emph{\bibinfo{journal}{Phys. Rev. B}} \textbf{\bibinfo{volume}{26}},
  \bibinfo{pages}{1352--1361} (\bibinfo{year}{1982}).

\bibitem{Pezze:2011ab}
\bibinfo{author}{{Pezz{\'e}}, L.} \emph{et~al.}
\newblock \bibinfo{title}{{Regimes of classical transport of cold gases in a
  two-dimensional anisotropic disorder}}.
\newblock \emph{\bibinfo{journal}{New Journal of Physics}}
  \textbf{\bibinfo{volume}{13}}, \bibinfo{pages}{095015}
  (\bibinfo{year}{2011}).

\bibitem{Lages:2000aa}
\bibinfo{author}{Lages, J.} \& \bibinfo{author}{Shepelyansky, D.~L.}
\newblock \bibinfo{title}{Cooper problem in the vicinity of the anderson
  transition}.
\newblock \emph{\bibinfo{journal}{Phys. Rev. B}} \textbf{\bibinfo{volume}{62}},
  \bibinfo{pages}{8665--8668} (\bibinfo{year}{2000}).

\bibitem{sherman_measurement_2012}
\bibinfo{author}{Sherman, D.}, \bibinfo{author}{Kopnov, G.},
  \bibinfo{author}{Shahar, D.} \& \bibinfo{author}{Frydman, A.}
\newblock \bibinfo{title}{Measurement of a superconducting energy gap in a
  homogeneously amorphous insulator}.
\newblock \emph{\bibinfo{journal}{Phys. Rev. Lett.}}
  \textbf{\bibinfo{volume}{108}}, \bibinfo{pages}{177006}
  (\bibinfo{year}{2012}).

\bibitem{dubi_theory_2006}
\bibinfo{author}{Dubi, Y.}, \bibinfo{author}{Meir, Y.} \&
  \bibinfo{author}{Avishai, Y.}
\newblock \bibinfo{title}{Theory of the magnetoresistance of disordered
  superconducting films}.
\newblock \emph{\bibinfo{journal}{Phys. Rev. B}} \textbf{\bibinfo{volume}{73}},
  \bibinfo{pages}{054509} (\bibinfo{year}{2006}).

\bibitem{PhysRevB.37.325}
\bibinfo{author}{Giamarchi, T.} \& \bibinfo{author}{Schulz, H.~J.}
\newblock \bibinfo{title}{Anderson localization and interactions in
  one-dimensional metals}.
\newblock \emph{\bibinfo{journal}{Phys. Rev. B}} \textbf{\bibinfo{volume}{37}},
  \bibinfo{pages}{325--340} (\bibinfo{year}{1988}).

\bibitem{fisher_boson_1989}
\bibinfo{author}{Fisher, M. P.~A.}, \bibinfo{author}{Weichman, P.~B.},
  \bibinfo{author}{Grinstein, G.} \& \bibinfo{author}{Fisher, D.~S.}
\newblock \bibinfo{title}{Boson localization and the superfluid-insulator
  transition}.
\newblock \emph{\bibinfo{journal}{Physical Review B}}
  \textbf{\bibinfo{volume}{40}}, \bibinfo{pages}{546--570}
  (\bibinfo{year}{1989}).

\bibitem{Feigelman:2010aa}
\bibinfo{author}{Feigel'man, M.~V.}, \bibinfo{author}{Ioffe, L.~B.},
  \bibinfo{author}{Kravtsov, V.~E.} \& \bibinfo{author}{Cuevas, E.}
\newblock \bibinfo{title}{Fractal superconductivity near localization
  threshold}.
\newblock \emph{\bibinfo{journal}{Annals of Physics}}
  \textbf{\bibinfo{volume}{325}}, \bibinfo{pages}{1390--1478}
  (\bibinfo{year}{2010}).

\bibitem{b_zimmermann_and_t_muller_and_j_meineke_and_t_esslinger_and_h_moritz_%
high-resolution_2011}
\bibinfo{author}{Zimmermann, B.}, \bibinfo{author}{M{\"u}ller, T.},
  \bibinfo{author}{Meineke, J.}, \bibinfo{author}{Esslinger, T.} \&
  \bibinfo{author}{Moritz, H.}
\newblock \bibinfo{title}{High-resolution imaging of ultracold fermions in
  microscopically tailored optical potentials}.
\newblock \emph{\bibinfo{journal}{New Journal of Physics}}
  \textbf{\bibinfo{volume}{13}}, \bibinfo{pages}{043007}
  (\bibinfo{year}{2011}).

\bibitem{petrov_interatomic_2001}
\bibinfo{author}{Petrov, D.~S.} \& \bibinfo{author}{Shlyapnikov, G.~V.}
\newblock \bibinfo{title}{Interatomic collisions in a tightly confined bose
  gas}.
\newblock \emph{\bibinfo{journal}{Physical Review A}}
  \textbf{\bibinfo{volume}{64}}, \bibinfo{pages}{012706}
  (\bibinfo{year}{2001}).

\end{thebibliography}

\begin{thebibliography}{10}

\bibitem{Zimmermann:2011aa}
{Zimmermann}, B., {M{\"u}ller}, T., {Meineke}, J., {Esslinger}, T., and
  {Moritz}, H.
\newblock {High-resolution imaging of ultracold fermions in microscopically
  tailored optical potentials}.
\newblock {\em New Journal of Physics}{ \bf 13}(4), 043007 (2011).

\bibitem{Goodman:2007aa}
Goodman, J. W. .~.
\newblock {\em Speckle phenomena in optics: theory and applications}.
\newblock Roberts and Company Publishers (2007).

\bibitem{Brantut:2012aa}
Brantut, J.-P., Meineke, J., Stadler, D., Krinner, S., and Esslinger, T.
\newblock Conduction of ultracold fermions through a mesoscopic channel.
\newblock {\em Science}{ \bf 337}(6098), 1069--1071 (2012).

\bibitem{Morsch:2006aa}
{Morsch}, O. and {Oberthaler}, M.
\newblock {Dynamics of Bose-Einstein condensates in optical lattices}.
\newblock {\em Reviews of Modern Physics}{ \bf 78}, 179--215 (2006).

\bibitem{Busch:1998aa}
Busch, T., Englert, B.-G., Rza{\.z}ewski, K., and Wilkens, M.
\newblock Two cold atoms in a harmonic trap.
\newblock {\em Foundations of Physics}{ \bf 28}, 549--559 (1998).

\bibitem{Petrov:2001aa}
Petrov, D.~S. and Shlyapnikov, G.~V.
\newblock Interatomic collisions in a tightly confined bose gas.
\newblock {\em Phys. Rev. A}{ \bf 64}(1), 012706 (2001).

\bibitem{Frohlich:2011aa}
Fr\"ohlich, B., Feld, M., Vogt, E., Koschorreck, M., Zwerger, W., and K\"ohl,
  M.
\newblock Radio-frequency spectroscopy of a strongly interacting
  two-dimensional fermi gas.
\newblock {\em Phys. Rev. Lett.}{ \bf 106}(10), 105301 (2011).

\bibitem{Sommer:2012aa}
Sommer, A.~T., Cheuk, L.~W., Ku, M. J.~H., Bakr, W.~S., and Zwierlein, M.~W.
\newblock Evolution of fermion pairing from three to two dimensions.
\newblock {\em Phys. Rev. Lett.}{ \bf 108}, 045302 (2012).

\bibitem{Giamarchi:1988aa}
Giamarchi, T. and Schulz, H.~J.
\newblock Anderson localization and interactions in one-dimensional metals.
\newblock {\em Phys. Rev. B}{ \bf 37}, 325--340 (1988).

\bibitem{Fisher:1989aa}
Fisher, M. P.~A., Weichman, P.~B., Grinstein, G., and Fisher, D.~S.
\newblock Boson localization and the superfluid-insulator transition.
\newblock {\em Phys. Rev. B}{ \bf 40}, 546--570 (1989).

\bibitem{Gimperlein:2005aa}
Gimperlein, H., Wessel, S., Schmiedmayer, J., and Santos, L.
\newblock Ultracold atoms in optical lattices with random on-site interactions.
\newblock {\em Phys. Rev. Lett.}{ \bf 95}, 170401 (2005).

\bibitem{Johnson:1998aa}
Johnson, C.~W., Bertsch, G.~F., and Dean, D.~J.
\newblock Orderly spectra from random interactions.
\newblock {\em Phys. Rev. Lett.}{ \bf 80}, 2749--2753 (1998).

\bibitem{Papenbrock:2002aa}
Papenbrock, T., Kaplan, L., and Bertsch, G.~F.
\newblock Odd-even binding effect from random two-body interactions.
\newblock {\em Phys. Rev. B}{ \bf 65}, 235120 (2002).

\bibitem{Jacquod:2000aa}
Jacquod, P. and Douglas~Stone, A.
\newblock Suppression of ground-state magnetization in finite-size systems due
  to off-diagonal interaction fluctuations.
\newblock {\em Phys. Rev. Lett.}{ \bf 84}, 3938--3941 (2000).

\bibitem{stadler_observing_2012}
Stadler, D., Krinner, S., Meineke, J., Brantut, J.-P., and Esslinger, T.
\newblock Observing the drop of resistance in the flow of a superfluid fermi
  gas.
\newblock {\em Nature}{ \bf 491}(7426), 736--739 (2012).

\end{thebibliography}

\end{document}